\documentclass[twocolumn,showpacs,amsmath,amssymb,aps,letterpaper]{revtex4}

\usepackage{amsmath}
\usepackage{graphicx,subfigure}
\usepackage{verbatim}

\newcommand{\lb}{{<}}
\newcommand{\rb}{{>}}
\newcommand{\bk}{Burridge-Knopoff}
\newcommand{\nn}{nearest-neighbor}
\newcommand{\lr}{long-range}
\newcommand{\st}{stress transfer}

\begin{document}

\title{Simulation of the \bk\ model of earthquakes with variable range \st}
\author{Junchao Xia}
\author{Harvey Gould}
\affiliation{Department of Physics, Clark University, Worcester, MA 01610}

\author{W. Klein}

\affiliation{Department of Physics and Center for Computational
Science, Boston University, Boston, MA 02215}
\author{J. B. Rundle}
\affiliation{Department of Physics and Center for Computational
Science and Engineering, University of California, Davis, CA 95616}

\begin{abstract}

Simple models of earthquake faults are important for understanding the
mechanisms for their observed behavior, such as Gutenberg-Richter scaling
and the relation between large and small events, which is the basis for
various forecasting methods. Although cellular automaton models have been
studied extensively in the \lr\ \st\ limit, this limit has not been studied
for the
\bk\ model, which includes more realistic friction forces and inertia. We
find that the latter model with
\lr\ \st\ exhibits qualitatively different behavior than both the
\lr\ cellular automaton models and the usual \bk\ model with nearest neighbor
springs, depending on the nature of the velocity-weakening friction force.
This results have important implications for our understanding of earthquakes
and other driven dissipative systems.

\end{abstract}

\maketitle

Earthquake faults are examples of driven dissipative
systems~\cite{geo}. Models of fault systems are important for understanding
scaling laws, the occurrence of characteristic
events, and the relation between small and large earthquakes~\cite{geo,
scholz, bow}. In addition to the benefits that would result from
understanding earthquake faults, understanding driven dissipative systems 
is important in physics and related areas.

Simulations of the \bk\ model~\cite{bk} have led to much insight.
The model consists of blocks connected by linear springs to their nearest
neighbors with spring constant
$k_c$. The blocks are also connected to a loader plate by linear springs
with spring constant
$k_{L}$, and rest on a surface with a nonlinear velocity-weakening stick-slip
friction force that depends on a parameter $\alpha$ which controls how
quickly the friction force decreases as the velocity is increased and
determines the amount of stress dissipated in an event. The model was
studied numerically in one dimension~\cite{bk} and more recently~\cite{cl,cl2,mori}.

An earthquake event is defined as a cluster of blocks that move due to the
initial slip of a single block. The moment of an event is 
$\propto
\sum_j u_j$, where the sum is over all the blocks in an event and
$u_j$ is the net displacement of block $j$. The main result of prior
studies~\cite{bk,cl,cl2} is that the moment distribution, $P(M)$, scales as
$M^{-b}$ for small localized events with an exponent $b
\approx 2$ for $\alpha > 1$. Large events show a pronounced peak in
$P(M)$ indicating a significant presence of characteristic or system-wide
events. In contrast, the number of earthquake events with
$s$ blocks, $P(s)$, does not show power law scaling. Note that the blocks are
connected only to their nearest neighbors in the
\bk\ model studied in Refs.~\cite{bk,cl,cl2,mori}.

In the cellular
automaton (CA) versions of the
\bk\ model~\cite{runbr,ofc} with \nn\ \st, $P(s)$ does not exhibit
power law scaling~\cite{grass}. A generalization of the CA models with \lr\
stress transfer yields considerable differences with the \nn\ CA
models~\cite{run1} and with the
\nn\ \bk\ model. In particular, the \lr\ CA models run in a
Gutenburg-Richter scaling mode and exhibit events of
all sizes, consistent with the system being near a mean-field critical
point~\cite{kl1, morein,run2,kl2}.
Quasi-periodic system-wide events are not
observed, and $P(s)$ exhibits power law scaling. Small and medium size
events can be interpreted as fluctuations about a free energy
minimum~\cite{kl1, run2}. Large events drive the system out of
``equilibrium'' from which the system decays back to an equilibrium
state~\cite{kl1}. We refer to this behavior as punctuated ergodicity, which 
also has been observed in the Southern California fault
system~\cite{tiamp1}. 

Both the CA and \nn\ \bk\ models lack several elements that would make them
more realistic representations of earthquake systems. In particular, the
latter does not include \lr\ stress transfer, and the
\lr\ CA models do not have inertia and more realistic friction laws.

In this letter we discuss the results of adding more realistic \lr\ stress
transfer to the \bk\ model. Our primary result is that the
statistical properties of this generalized \bk\
model depend not only on the range $R$ of the stress transfer, but also on
the characteristics of the friction force. For large
velocity weakening, that is, large $\alpha$ (see
Eq.~\eqref{Burridge-Knopoff2}), the properties of the \bk\ model are
largely independent of 
$R$, although there are some characteristics of the
model that differ from the
\nn\ ($R=1$) model and are reminiscent of the \lr\
CA models. For small $\alpha$ and large $R$, the results
are identical to the \lr\ CA models.

The usual ($R=1$) \bk\ model in one dimension is governed by the
equation of motion
\begin{equation}
\label{Burridge-Knopoff1}
m \ddot x_j = k_c (x_{j+1} - 2x_j + x_{j-1}) -
k_{L}x_j - F(v + {\dot x_j}),
\end{equation}
where $x_j$ is the displacement of block $j$ from its equilibrium position,
$v$ is the speed of the substrate moving to the left with a fixed loader
plate,
$F(\dot x) = F_0\phi(\dot x/\tilde v)$ is the velocity-dependent friction
force,
$\tilde v$ is a characteristic velocity, and $m$ is the mass of a block.

As in Ref.~\cite{cl} we introduce scaled variables $\tau = \omega_{p}t$, 
$\omega^{2}_{p} = k_{L}/m$, and $u_j
= (k_{L}/F_0)x_j$, and rewrite Eq.~(\ref{Burridge-Knopoff1}) as
\begin{equation}
\label{Burridge-Knopoff2}
\ddot u_j = \ell^{2}(u_{j + 1} -2u_j + u_{j - 1}) - u_j -
\phi(2\alpha \nu + 2\alpha{\dot u}_j),
\end{equation}
with $2\alpha =
\omega_{p}F_0/k_{L}\tilde v$, $\ell^{2} =
k_c/k_{L}$, and $\nu = v k_{L}/(\omega_{p}F_0)$; the dots now denote
differentiation with respect to $\tau$. The friction force is
\begin{equation}
\label{Burridge-Knopoff3}
\phi(y)=
\begin{cases}
(-\infty, 1], & \text{$y = 0$}\\
{1 - \sigma\over 1 + {y\over 1 - \sigma}}, & \text{$y > 0$}.
\end{cases}
\end{equation}

We generalize the \bk\ model by assuming that a block is connected to 
$R$ neighbors (in each direction) with spring constant $k_c/R$;
$R=1$ corresponds to the usual \bk\ model. We use the second- and
fourth-order Runge-Kutta algorithms with time step $\Delta t = 0.001$ to
solve 
Eq.~\eqref{Burridge-Knopoff2}, modified for arbitrary $R$, with the friction
force given in Eq.~\eqref{Burridge-Knopoff3}. Both algorithms and other
fourth-order algorithms give similar results. 

A block is defined to be stuck if its speed is less than the parameter
$v_0$, its velocity is decreasing, and the stress (the force
due to the springs coupled to it) is smaller than the maximum static friction
force $F_0$ (taken to be unity). If a block is stuck, its velocity is set equal to
zero, and the static friction force cancels the stress.
A stuck block will accelerate if the stress is greater than $F_0$; in this case the friction force drops to the
value 
$1-\sigma$. An earthquake begins with the slip of a block and ends when all
blocks become stuck. Blocks can become stuck and then slip again during an
event.
The results in this paper are
for $\ell = 10$, $\sigma=0.01$, $v_0=10^{-5}$, $N=5000$ blocks, $R=1$, 100, 500, 1000, $\alpha=0.0$, 0.5, and 2.5, and $10^6$
events.

We initially set $\dot u_j = 0$ for all $j$ and assign random displacements
to all the blocks. We compute the force on all the blocks and update $\dot
u_j$ and $u_j$ for all $j$. We
continue these updates until all blocks become stuck. We then move the
substrate (the loader plate is fixed) until the stress on one block
exceeds unity. This stress loading mechanism is
known as the ``zero velocity limit''~\cite{cl2} and is equivalent to setting
$\nu=0$.

\begin{figure}[tp]
\begin{center}
\subfigure[\label{fig:PMa} {$\alpha = 2.5$.}]{
\includegraphics[scale=0.56]{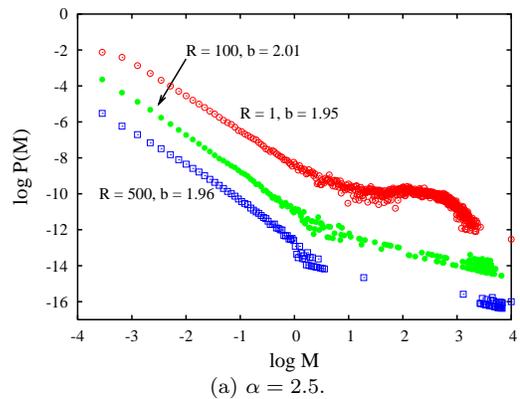}}
\subfigure[\label{fig:PMb} {$\alpha = 0.5$.}]{
\includegraphics[scale=0.56]{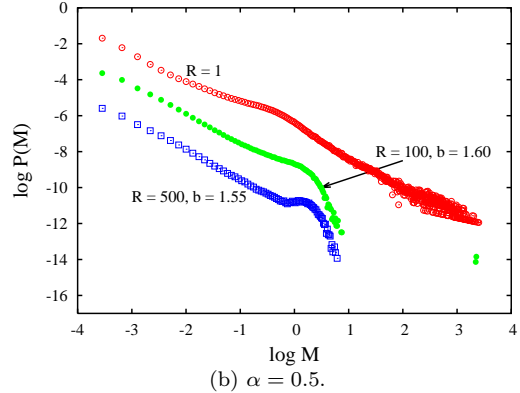}}
\subfigure[\label{fig:PMc} {$\alpha = 0$.}]{
\includegraphics[scale=0.56]{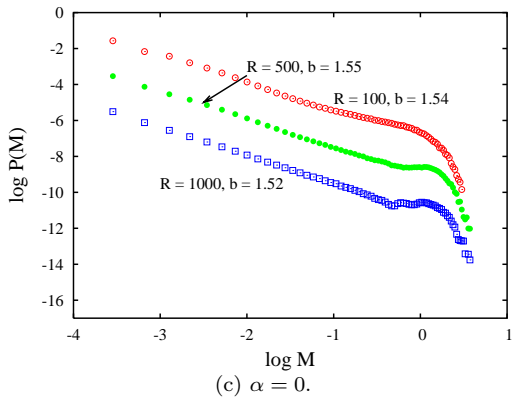}}
\vspace{-1pc}
\caption{\label{fig:PM}Distribution of the moment $M$ of the events
$P(M)$.
For $\alpha = 2.5$ and $R=1$ the slope $b \approx
2$, consistent with
Ref.~\cite{cl}; $b$ does not
change significantly as $R$ increases. For $\alpha=0.5$ the scaling range
becomes better defined as
$R$ increases and the slope converges to $b \approx 1.5$. For
$\alpha=0$, $b\approx 1.5$ for $R \gtrsim 100$; there is no power
law behavior for $R=1$ and small $\alpha$. All plots have been displaced
vertically for clarity.}
\end{center}
\vspace{-2pc}
\end{figure}

\begin{figure}[tp]
\begin{center}
\subfigure[\label{fig:eventsa} {$\alpha = 2.5$.}]{
\includegraphics[scale=0.56]{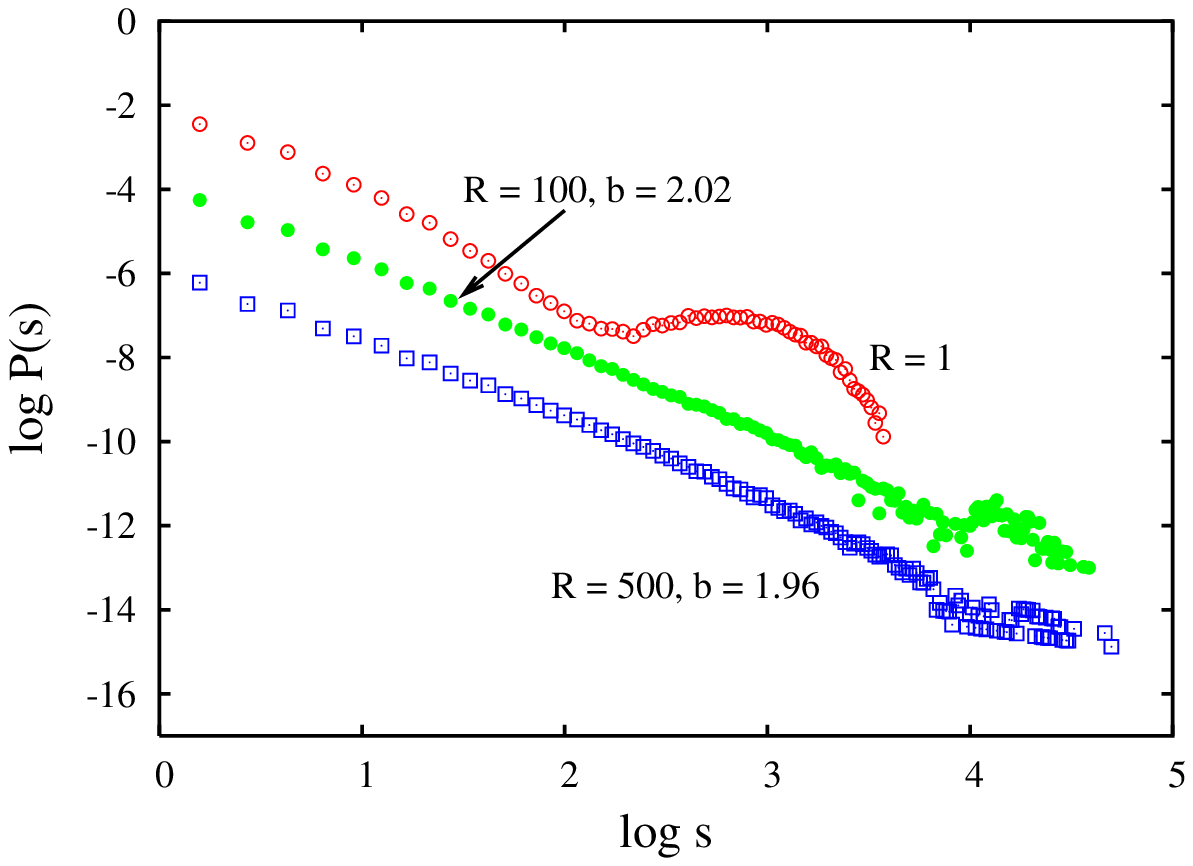}}
\subfigure[\label{fig:eventsc} {$\alpha = 0$.}]{
\includegraphics[scale=0.56]{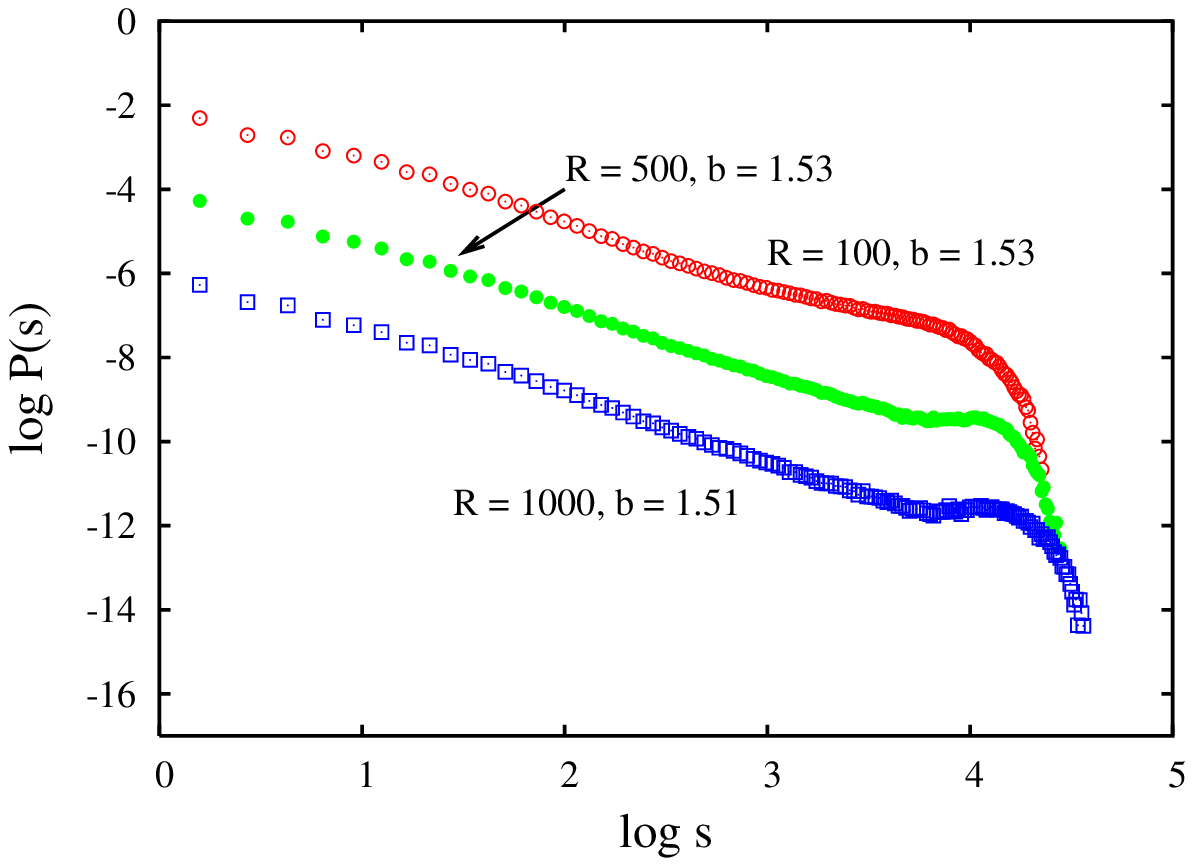}}
\vspace{-1pc}
\caption{\label{fig:events}Distribution of events of size $s$, $P(s)$.
$P(s)$ does not exhibit power law behavior for
$\alpha=2.5$ and $R=1$,
but does so for larger $R$ with $b \approx 2$. For
$\alpha = 0$, $P(s) \sim s^{-b}$ with $b \approx 1.5$ for $R \gtrsim 100$. Note
that $s$ can exceed
$N$ because a block can fail multiple times during an event.}
\end{center}
\vspace{-2.0pc}
\end{figure}

In Fig.~\ref{fig:PM} we plot $P(M)$ for various
values of
$R$ and $\alpha$. For $\alpha = 2.5$ and $R = 1$, $P(M) \sim M^{-b}$ with $b
\approx 2$ as in Ref.~\cite{cl}. For larger values of
$R$, $b$ remains $\approx 2$, but the scaling range decreases and the
characteristic earthquakes ($M>1$) become better defined. 
For $\alpha = 0.5$ and $R \gtrsim 100$, $b$ appears to approach 1.5 as
$R$ increases, and there are fewer
characteristic earthquakes (see Fig.~\ref{fig:PMb}). The results in
Fig.~\ref{fig:PMc} for $\alpha=0$ are consistent with the \lr\ CA
models~\cite{kl1,morein} for which $b=3/2$. 
The distribution of events, $P(s)$, does not exhibit power law behavior for
$R=1$ and $\alpha=2.5$, consistent with Ref.~\cite{cl},
but does so for larger $R$ with $P(s) \sim s^{-b}$ and $b \approx 2$ (see
Fig.~\ref{fig:events}). For small
$\alpha$, $b \approx 1.5$ for $R \gtrsim 100$.

\begin{figure}[tp]
\begin{center}
\subfigure{
\includegraphics[scale=0.31]{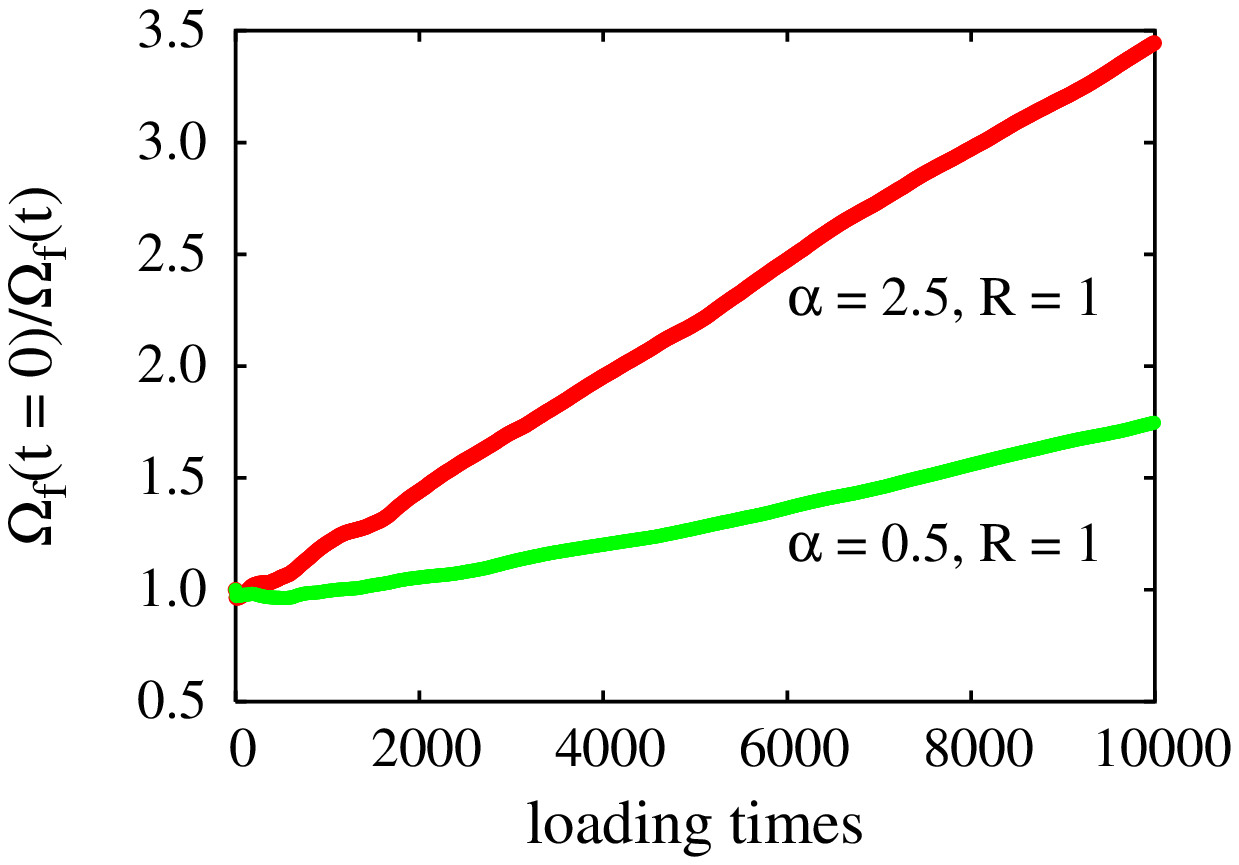}}
\subfigure{
\includegraphics[scale=0.31]{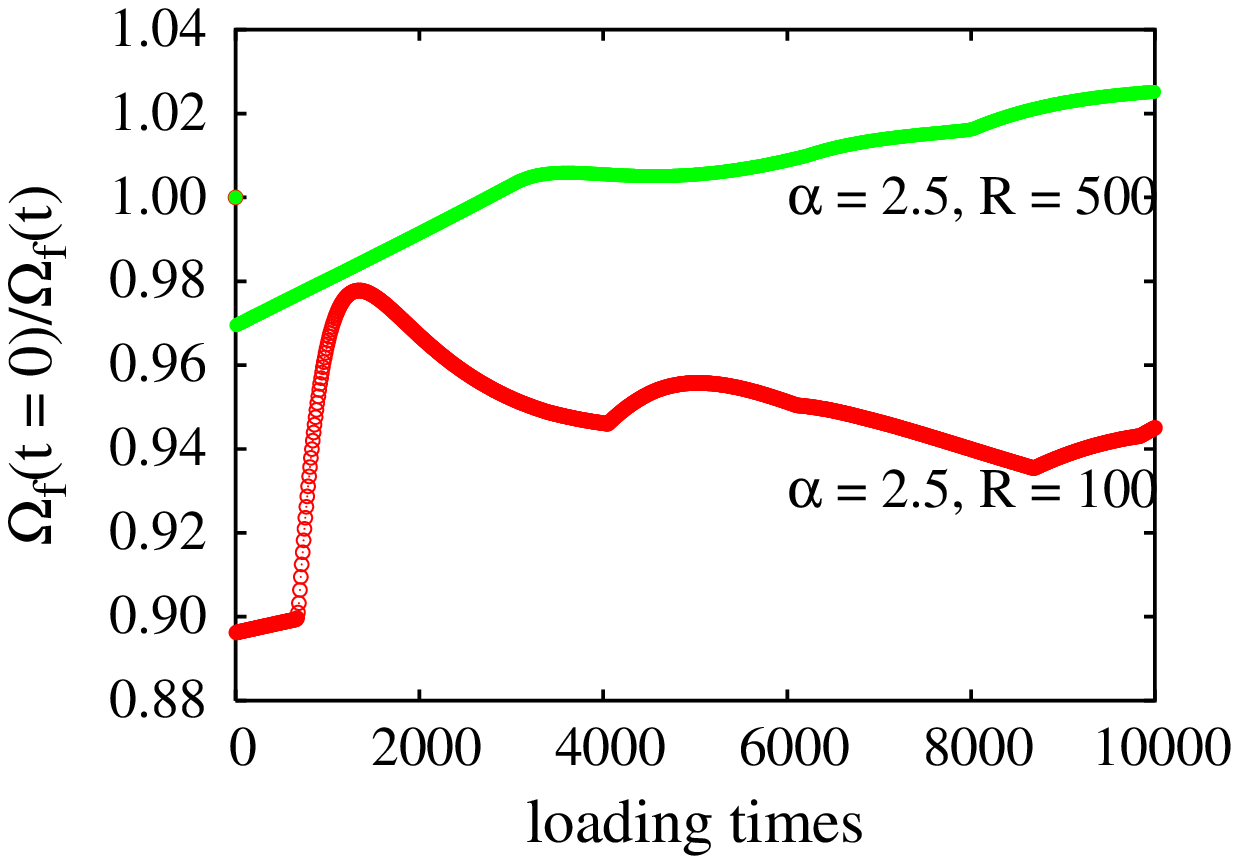}}
\subfigure{
\includegraphics[scale=0.31]{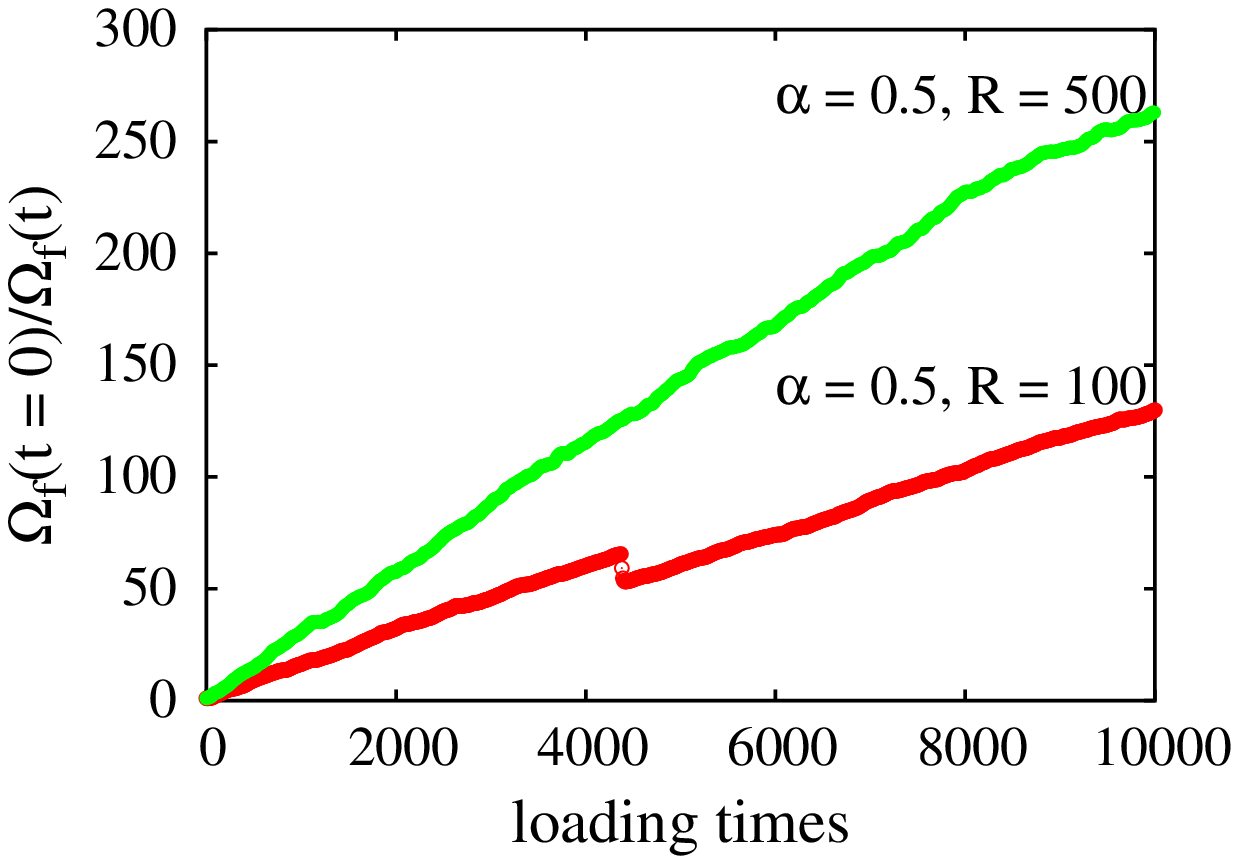}}
\subfigure{
\includegraphics[scale=0.31]{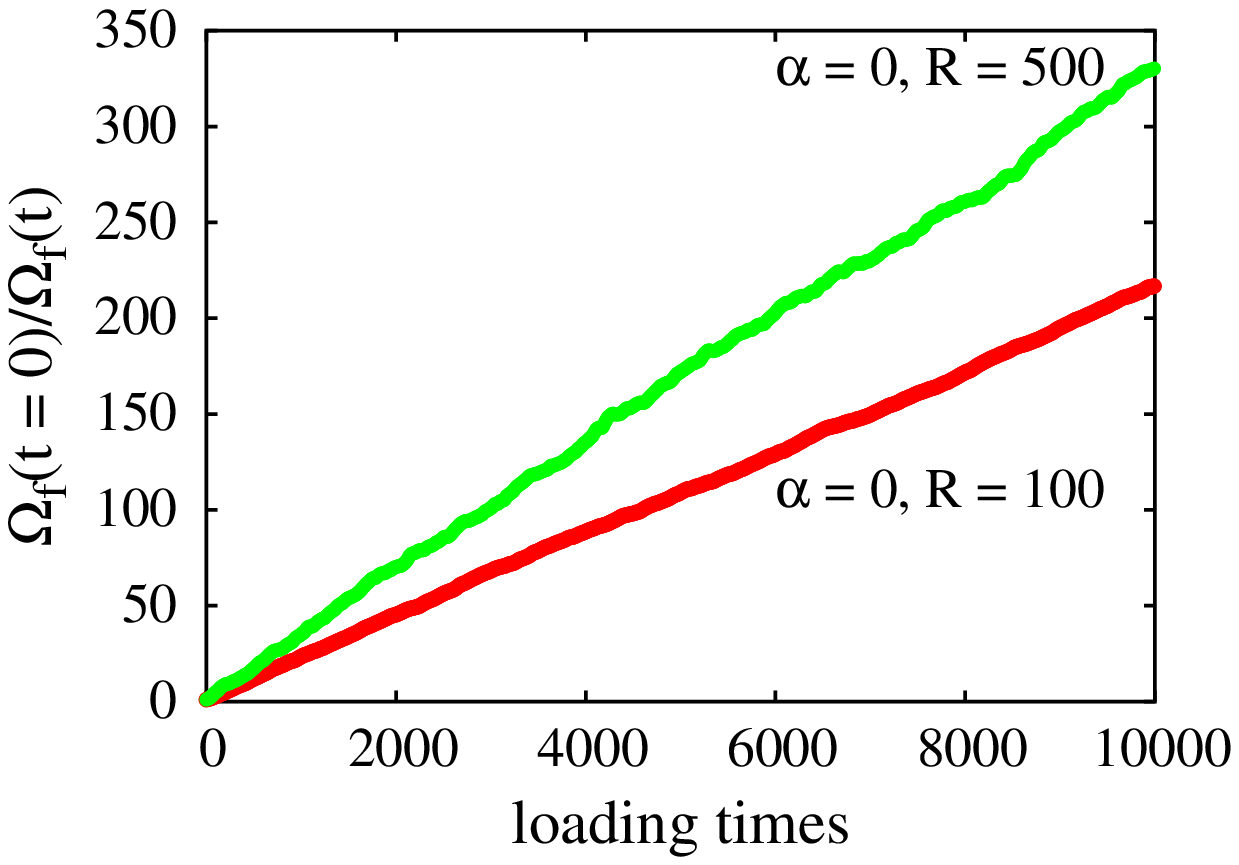}}
\vspace{-1pc}
\caption{\label{fig:stressmetric}The inverse stress metric versus the number of loading times. Note the different
vertical scales. The system appears ergodic for $\alpha=2.5$ and $R=1$,
but is nonergodic for larger values of $R$. For
$\alpha=0.5$, the slope of $\Omega_f^{-1}(t)$ becomes larger as $R$ increases.
For
$\alpha=0.5$ and
$R=100$, the system exhibits punctuated ergodicity similar to the behavior
of the \lr\ CA model.}
\end{center}
\vspace{-2.0pc}
\end{figure}

We now consider the existence of ergodicity and determine the
metric $\Omega_f(t)$~\cite{TM}. We take $f_j(t)$ to be a quantity associated with
block
$j$ and define
\begin{align}
\label{tm1}
{\overline f}_j(t) &= {1\over
t}\!\int_{0}^{t}\! f_j(t^{\prime})\,dt^{\prime} \\
\label{tm2}
\lb f(t) \rb &= {1\over N} \sum_{j = 1}^{N}f_j(t)\\
\label{tm3}
\Omega_{f}(t)&= {1\over N}\sum_{j = 1}^{N}\big\lbrack {\overline f}_j(t)
-
\lb f(t) \rb\big\rbrack^{2}.
\end{align}
If the system is ergodic, 
$\Omega_{f}(t)\propto 1/t$~\cite{TM}. The metric studied in the CA models is
the stress metric. In Fig.~\ref{fig:stressmetric} we show the
inverse stress metric $\Omega_{f}^{-1}(t)$ for the \bk\ model as a function
of the number of times the substrate is moved (loading times). For $R=1$ the system appears to be
ergodic, unlike the short-range CA models~\cite{kleinergodic}.
For
$\alpha=2.5$, the system ceases to be ergodic for larger values of $R$. In
contrast, for
$\alpha=0.5$ the system remains ergodic, and the slope of $\Omega_f^{-1}$
becomes larger as $R$ increases. In particular, for $\alpha \leq 0.5$ and
large
$R$, the behavior of
$\Omega_f$ is similar to the stress metric in the
\lr\ CA models~\cite{ferg}.

\begin{figure}[tp]
\begin{center}
\includegraphics[scale=0.6]{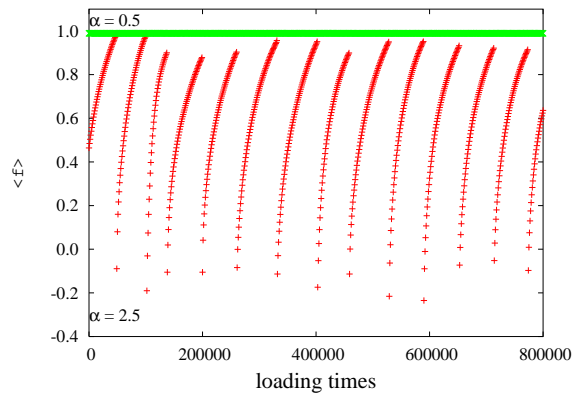}
\vspace{-0.5pc}
\caption{\label{fig:meanstress}The mean
stress $\lb f \rb$ as a function of the number of loading times for $R=500$. Note the
fluctuating behavior for $\alpha=0.5$ and the quasiperiodic behavior for
$\alpha=2.5$.}
\end{center}
\vspace{-2pc}
\end{figure}

To understand this behavior, we plot the time series of the
mean stress $\lb f \rb$ per block just after an event for $R = 500$ (see Fig.~\ref{fig:meanstress}). For
$\alpha=2.5$ $\lb f(t) \rb$ is quasi-periodic and ranges from
$\approx -0.2$ to almost 1.0 with a mean of $\approx 0.4$. That is, small earthquakes accumulate stress
locally and characteristic earthquakes release the stress globally and
quasi-periodically. The periodicity of characteristic earthquakes for
$R=1$ and $\alpha=2.5$ was also analyzed in Ref.~\cite{mori}. The distribution
of characteristic earthquakes and other results~\cite{jun2} show that
the periodicity becomes better defined as $R$ is increased. The
quasi-periodic behavior in Fig.~\ref{fig:meanstress} is reminiscent of the
stress versus time curves observed in laboratory experiments with
rocks~\cite{km}. In contrast, for
$\alpha = 0.5$, $\lb f(t) \rb$ fluctuates about
$\approx 0.99$ (see Fig.~\ref{fig:meanstress}). The latter behavior
is observed in \lr\ CA models~\cite{ferg}.

\begin{figure}[tp]
\begin{center}
\includegraphics[scale=0.58]{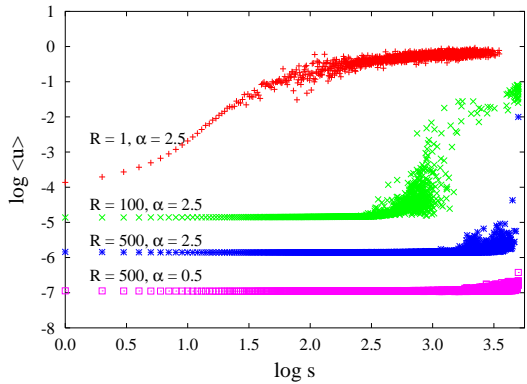}
\vspace{-0.5pc}
\caption{\label{fig7}The mean displacement $\lb u \rb$ (slip) of the
blocks as a function of the number of blocks $s$ in an event. (Each block is
counted once even if it fails multiple times.)}
\end{center}
\vspace{-2.0pc}
\end{figure}

Other statistical measures of the \lr\ \bk\ model approach those of the
\lr\ CA model. For example, the mean
displacement (slip) of the blocks during an event, $\lb u \rb$, as a
function of
$s$, the number of blocks in an event, is shown in Fig.~\ref{fig7}. For large
$R$ the mean displacement is independent of the size of an event up to the largest events, as in the \lr\ CA models. The range of
events over which the mean slip is a constant increases with $R$.

In summary, we have found two possible scaling regimes, one of which ($\alpha
\rightarrow 0$ and $R>>1$), is associated with an equilibrium
spinodal critical point with an exponent $b = 3/2$ consistent
with the \lr\ CA models~\cite{kl1,morein,run2,kl2,ferg}. The power law scaling of the
moment for 
$\alpha = 2.5$ and $R=1$ with $b \approx 2$ has been associated with
critical behavior~\cite{cl,cl2}. This latter behavior and the
ergodicity of the stress metric for this case suggests the presence of
another critical point for short-range systems, but this interpretation needs more
investigation. That is, the
\bk\ model appears to have two scaling regimes with qualitatively
different behavior.

The
apparent dependence of
$b$ on $\alpha$ suggests that larger system sizes as well as longer
run times should be investigated. However, because real faults are finite
and the number of events observed is small, the
$\alpha$-dependence of $b$ seen in the \lr\ \bk\ model may accurately reflect
the behavior of real faults.

For $\alpha=2.5$ and large $R$, our results resemble those observed in
laboratory experiments on rocks. As the range
$R$ is increased and $\alpha$ is made smaller, our results more closely
resemble the
\lr\ CA models. This wide range of behavior indicates that the physics
of several models of earthquake faults~\cite{bk,cl,runbr,ofc} can be obtained
from the generalized \bk\ model with the appropriate choice of $R$ and
$\alpha$.

Our results imply that real earthquake faults and laboratory rocks
can have different statistical distributions of events and different physical
characteristics due to the details of the friction force as well as range of
the stress transfer. This dependence means that we must develop ways of
determining the friction force with considerably more accuracy if we want to
understand the relation between physical processes and observed earthquake
phenomena, and if we are to be able to predict earthquakes or even determine
if they are predictable. 

\vspace{1pc}
\begin{acknowledgements}
This work has been supported in part by DOE grant DE-FG02-95ER14498 (WK and
JX) and DE-FG02-04ER15568 (JBR). The simulations were done at Clark with the partial support of NSF grant
DBI-0320875.
\end{acknowledgements}

\end{document}